\documentclass{elsart}

\usepackage{graphicx}
\usepackage{amssymb}

\usepackage[square,comma,sort&compress]{natbib}

\def\pT{p_{{}_\mathrm{T}}}

\begin{document}

\begin{frontmatter}

\title{%
    Upper limit on the decay \boldmath$\Sigma(1385)^-\to\Sigma^-\gamma$,
    and cross section for \boldmath$\gamma\Sigma^-\to\Lambda\pi^-$
}

The SELEX Collaboration
\author[Protvino]{V.V.~Molchanov\corauthref{cor}},
\corauth[cor]{Corresponding author.}
\ead{molchanov@mx.ihep.su}
\author[PNPI]{G.~Alkhazov},
\author[PNPI]{A.G.~Atamantchouk\thanksref{tra}},
\author[ITEP]{M.Y.~Balatz\thanksref{tra}},
\author[PNPI]{N.F.~Bondar},
\author[Rochester]{D.~Casey},
\author[Fermi]{P.S.~Cooper},
\author[Flint]{L.J.~Dauwe},
\author[ITEP]{G.V.~Davidenko},
\author[MPI]{U.~Dersch\thanksref{trb}},
\author[ITEP]{A.G.~Dolgolenko},
% ...
\address[Fermi]{Fermi National Accelerator Laboratory, Batavia, IL 60510, U.S.A.}
\address[Protvino]{Institute for High Energy Physics, Protvino, Russia}
% ...
\address[ITEP]{Institute of Theoretical and Experimental Physics, Moscow, Russia}
\address[MPI]{Max-Planck-Institut f\"ur Kernphysik, 69117 Heidelberg, Germany}
% ...
\address[PNPI]{Petersburg Nuclear Physics Institute, St. Petersburg, Russia}
% ...
\address[Flint]{University of Michigan-Flint, Flint, MI 48502, U.S.A.}
\address[Rochester]{University of Rochester, Rochester, NY 14627, U.S.A.}
% ...
\thanks[tra]{deceased}
\thanks[trb]{Present address: Infinion, M\"unchen, Germany}
\thanks[trc]{Present address: University of California at Irvine, Irvine, CA 92697, USA}
\thanks[trd]{Present address: Instituto de F\'{\i}sica da Universidade Estadual de Campinas, UNICAMP, SP, Brazil}
\thanks[tre]{Present address: Kafkas University, Kars, Turkey}
\thanks[trf]{Present address: Physik-Department, Technische Universit\"at M\"unchen, 85748 Garching, Germany}
\thanks[trg]{Present address: The Boston Consulting Group, M\"unchen, Germany}
\thanks[trh]{Present address: Lucent Technologies, Naperville, IL}
\thanks[tri]{Present address: SPSS Inc., Chicago, IL}
\thanks[trj]{Present address: University of Alabama at Birmingham, Birmingham, AL 35294}
\thanks[trk]{Present address: Imadent Ltd.,\ Rehovot 76702, Israel}
\thanks[trl]{Present address: S\"uleyman Demirel Universitesi, Isparta, Turkey}
\thanks[trm]{Present address: DOE, Germantown, MD}
\thanks[trn]{Present address: Solidum, Ottawa, Ontario, Canada}
\thanks[tro]{ Present address: Siemens Medizintechnik, Erlangen, Germany}
\thanks[trp]{Present address: Allianz Insurance Group IT, M\"unchen, Germany}

\begin{abstract}

Coherent $\Lambda\pi^-$ production on Pb of
$600\,\mathrm{GeV}$ $\Sigma^-$ hyperons
has been studied
with the SELEX facility at Fermilab.
Using the Primakoff formalism,
we set a $90\%$ CL upper limit on the radiative decay width
$\Gamma[\Sigma(1385)^-\to\Sigma^-\gamma]<9.5\,\mathrm{keV}$,
and estimate the cross section for
$\gamma\Sigma^-\to\Lambda\pi^-$
at $\sqrt{s}\approx1.385\,\mathrm{GeV}$
to be~$56\pm16\,\mu\mathrm{b}$.
\end{abstract}


\end{frontmatter}

\section{Introduction}

Radiative decays of hadrons and other electromagnetic processes play
an important role in the phenomenology of high energy physics.
Interactions of real and virtual photons with electric charges
of quark fields provide unique information about the quark content
of hadrons and about phenomenological parameters such as form factors,
magnetic moments, polarizabilities,~etc.
Being simpler to analyze than purely hadronic phenomena,
such processes serve as a good testing ground for models of hadrons.

A lot of data is available on radiative decays
of mesons~\cite{review-mesons}.
Electromagnetic properties of the $N$ and $\Delta$ baryons
have also been studied extensively in resonance photoproduction
and electroproduction reactions~\cite{review-photoelectro}.
However, experimental data on radiative decays of hyperons
(we do not include {\em weak} radiative decays)
are rather scarce~\cite{Landsberg:1996gb}.
Only $\Sigma^0\to\Lambda\gamma$ and $\Lambda(1520)\to\Lambda\gamma$
decays have been measured thus far. There are also indirect
model-dependent estimates for
$\Lambda(1405)\to\Lambda/\Sigma^0\,\gamma$ decays,
based on the study of radiative capture of kaons in $K^-$-mesonic atoms.
No other radiative decays of hyperons have been observed,
and previous upper limits are not very
sensitive~\cite{Arik:1977mt,Colas:1975ck}.

Electromagnetic decays
$\Sigma(1385)^-\to\Sigma^-\gamma$ and $\Xi(1530)^-\to\Xi^-\gamma$
are of particular interest. In the limit of unbroken SU(3) symmetry,
these decays are forbidden by $U$-spin conservation~\cite{Lipkin:1973rw}.
Consequently, their observation
could provide information on the SU(3) breaking mechanism.
The decay $\Sigma(1385)^-\to\Sigma^-\gamma$ has been considered
in the framework of SU(3) and SU(6) symmetry-breaking
schemes~\cite{Lipkin:1973rw,Lipkin:1992vw,Akhiezer-eng},
in the nonrelativistic quark model of Isgur and Karl~\cite{Darewych:1983yw},
the MIT bag model~\cite{Hackman:1978am},
Skyrme model~\cite{Schat:1995mt,Abada:1996db,Haberichter:1997cp},
heavy baryon chiral perturbation theory~\cite{Butler:1993pn,Napsuciale:1997ny},
quenched lattice calculations~\cite{Leinweber:1993pv},
and in other treatments~\cite{Sahoo:1995kg,Wagner:1998bu,Wagner:2000ii}.
Predictions for the radiative width vary from~1 to~$9.5\,\mathrm{keV}$,
although most models predict a width of less than $4\,\mathrm{keV}$.

The direct decay of $\Sigma(1385)^-\to\Sigma^-\gamma$
is difficult to measure because of its small partial width and
high background from the $\Sigma(1385)^-\to\Sigma^-\pi^0$ decay
with one lost photon. However, the width for this
and many other decays of the type $a \to h+\gamma$
can be measured in the inverse reaction
of coherent electromagnetic production
in the Coulomb field of atomic nuclei
\begin{equation}
\label{re-coulomb}
h + (A,Z) \to a + (A,Z)
\end{equation}
The cross section for such reactions is proportional
to the radiative decay width $\Gamma(a \to h+\gamma)$.
Thus $\Gamma(a \to h+\gamma)$ can be determined by
measuring the absolute cross section of the Coulomb
contribution to Reaction~(\ref{re-coulomb}).
This method was proposed initially by Primakoff, Pomeranchuk and
Shmushkevich~\cite{primakoff-1951-prim,pomeranchuk-1961-prim},
and is usually referred to as the Primakoff effect.
Detailed description of the method may be found
in many review papers~\cite{review}
and references therein.

In this Letter, we present results of a search
for the Coulomb production reaction
\begin{equation}
\label{re-main}
\begin{array}{rcl}
\Sigma^- + \mathrm{Pb} & \to & \Sigma(1385)^- + \mathrm{Pb}\\
& & \kern0.36em
    \hbox{\vrule height3.0ex depth-0.6ex}
    \kern-0.37em
    \to \Lambda\pi^-
\end{array}
\end{equation}
at a beam energy of approximately $600\,\mathrm{GeV}$ in an experiment
using the SELEX spectrometer (E781) at Fermilab.

\section{Experimental apparatus}

The SELEX facility~\cite{Smith:1997ud}
is a forward magnetic spectrometer
with scintillation counters and hodoscopes, proportional
and drift chambers, silicon microstrip beam and vertex detectors,
additional downstream microstrip stations in the beam region,
three lead glass photon detectors, a hadron calorimeter,
two transition radiation detectors~(TRD),
and a multiparticle RICH counter.

The experiment was designed mainly to study
production and decays of charm baryons in a hyperon
beam~\cite{russ-physics}.
It emphasized the forward production
($x_{{}_\mathrm{F}}>0.1$) region and, consequently, had high
acceptance for exclusive low multiplicity processes.
Studies of Coulomb production were performed in parallel
with the main charm-physics program and several other measurements.
This imposed certain limitations
on the trigger, geometry, choice of targets and consequent statistics.

\begin{figure}[!t]
\centering
\includegraphics[width=\hsize]{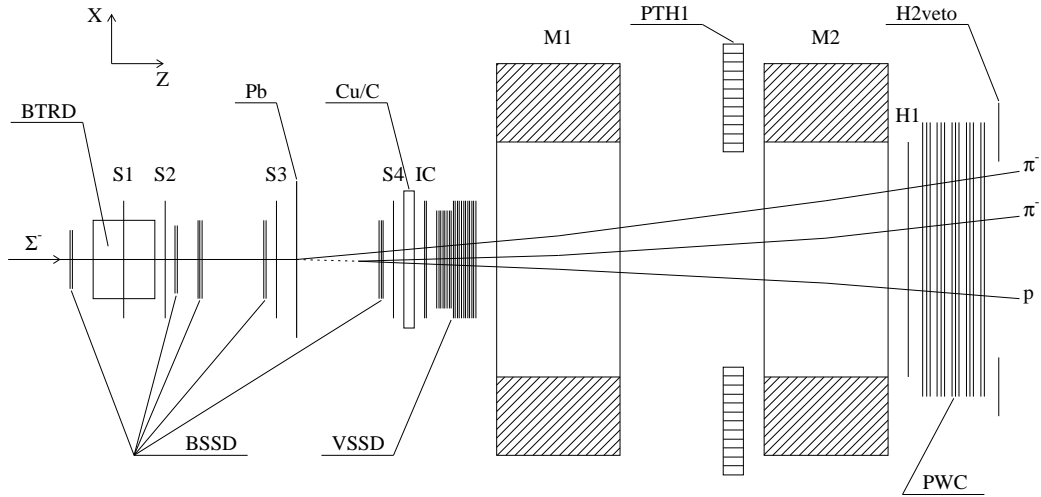}
\caption{%
    Schematic plan view (not to scale) of the SELEX apparatus.
    Shown are major detectors used for these measurements:
    BTRD~--- beam transition radiation detector;
    BSSD~--- beam silicon strip detector;
    VSSD~--- vertex silicon strip detectors;
    PWC~--- proportional wire chambers;
    M1, M2~--- analyzing magnets;
    Pb~--- target for this measurement;
    Cu/C~--- targets for charm production measurements;
    S1--S4, IC~--- scintillation counters;
    H1, H2veto~--- trigger hodoscope and veto;
    PHT1~--- lead glass photon detector.
}
\label{fig-setup_s1385}
\end{figure}

We report studies based on a special short exposure
with a thin ($1.98\,\mathrm{mm}$) Pb target.
Figure~\ref{fig-setup_s1385} shows a portion of the SELEX apparatus
relevant for this measurement.
The origin of the coordinate system was chosen to be
in the middle of the downstream surface of the downstream target,
and the $Z$-axis was directed along the beam.
The Pb target was located at $Z=-55\,\mathrm{cm}$,
while standard targets for charm measurements
were removed for most of the exposure.
%
%
At the position of the Pb target,
the beam was composed of ${\approx}50\%\;\pi^-$, ${\approx}48\%\;\Sigma^-$
and ${\approx}1\%\;\Xi^-$.
Fractions of $\bar p$ and $\Omega^-$ were estimated to be less than~$0.1\%$.
The beam transition-radiation detector provided reliable separation
of baryons from~$\pi^-$.
Silicon strip detectors (most of which had $4\,\mu\mathrm{m}$
transverse position resolution) provided parameters of the beam
and secondary tracks in the target region. After deflection by
analyzing magnets, tracks were measured in 14 planes
of $2\,\mathrm{mm}$ proportional wire chambers.
The absolute momentum scale was calibrated
using the $K^0_\mathrm{S}\to\pi^+\pi^-$ decays.
Mass resolution in the $\Sigma(1385)$ region
was measured using the beam $\Xi^-\to\Lambda\pi^-$ decays,
and was found to be~$\approx6\,\mathrm{MeV}$.

Reaction~(\ref{re-main})
was singled out with the help of a special exclusive trigger.
This trigger used scintillation counters S1--S4 to define beam time,
and trigger hodoscope H1 to require three charged tracks.
Pulse heights in the scintillation counters IC
were used to select three charged tracks,
while pulse heights in counters S2 and S3 were used
to suppress interactions initiated upstream of the Pb target.
Thus, the trigger limited the decay volume of the $\Lambda$
to the region between the S3 and IC counters,
or from $-68\,\mathrm{cm}$ to $2.5\,\mathrm{cm}$ in SELEX coordinates.
Finally, to reduce the background trigger rate to an acceptable level,
the aperture was limited by veto counters H2veto,
which had very little effect on the efficiency for Reaction~(\ref{re-main}).

A special on-line filter was used
to reduce the number of exclusive events written to tape.
This selected events that had at least
two reconstructed segments downstream of the analyzing magnets.
Very loose criteria were imposed on the number of hits
in the tracking detectors to reduce processing time.
All these requirements were not very restrictive, and are expected
to have only minor effect on the process of interest.

\section{Data analysis}

Events for Reaction~(\ref{re-main})
were selected by requiring a reconstructed beam track and
three charged tracks in the final state,
of which one was positive and two negative.
The beam particle had to be identified as a $\Sigma^-$ ($\Xi^-$)
by the beam transition-radiation detector.
There was no explicit identification
of the produced particles:
a proton mass was assigned to the positive track
and a pion mass to each negative track.
It was then required that one of the pions and the proton
formed a good vertex (referred to as secondary)
with an effective mass close to the $\Lambda$ mass:
$1.11\,\mathrm{GeV}<M(p\pi^-)<1.122\,\mathrm{GeV}$.
Then, a primary vertex was formed using
the remaining pion, the beam, and the reconstructed $\Lambda$ tracks.
Typical resolution for vertices along the beam direction~($Z$)
was~$1\,\mathrm{cm}$.
To suppress inclusive ($\Lambda\pi^-+X$) background,
the energy sum of the observed particles
was required to be within $\pm20\,\mathrm{GeV}$
(${\approx}$~${\pm}$~two standard deviations) of the beam energy.
For further supression of these events,
the most upstream photon detector~(PHT1) was used as a guard system,
requiring that any registered energy be less than $2\,\mathrm{GeV}$.

\begin{figure}[!b]
\begin{minipage}[t]{0.47\hsize}
\centering
\includegraphics[width=\hsize]{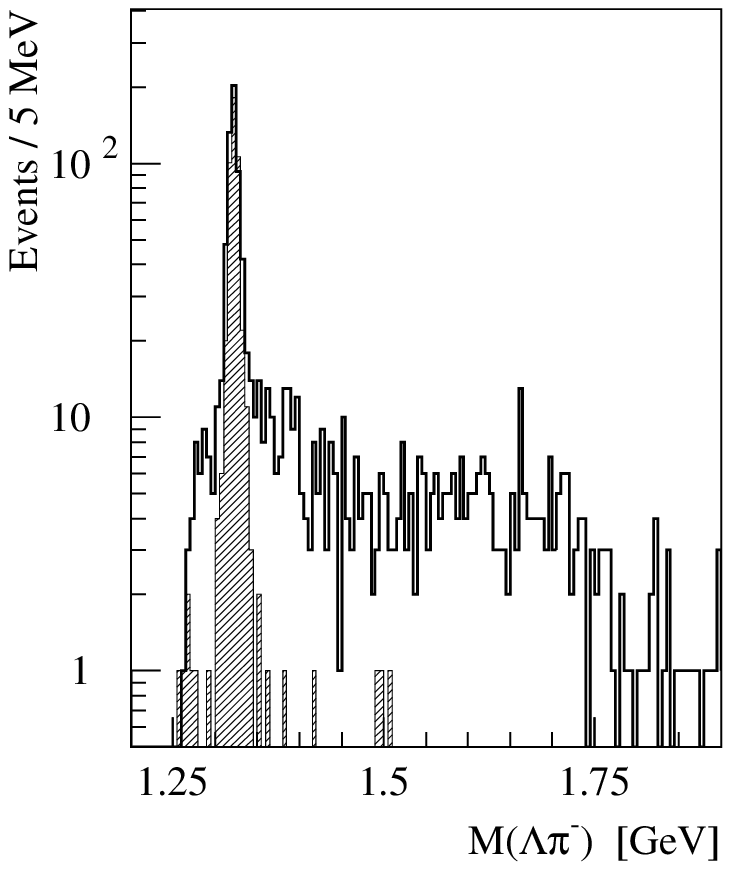}
\caption{%
    $M(\Lambda\pi^-)$ distributions for $\pT^2<0.001\,\mathrm{GeV}^2$.
    Events with a primary vertex in the target region are shown
    by the solid line, and events corresponding to sidebands
    are shown hatched.
}
\label{fig-bgr}
\end{minipage}
\hfill
\begin{minipage}[t]{0.47\hsize}
\centering
\includegraphics[width=\hsize]{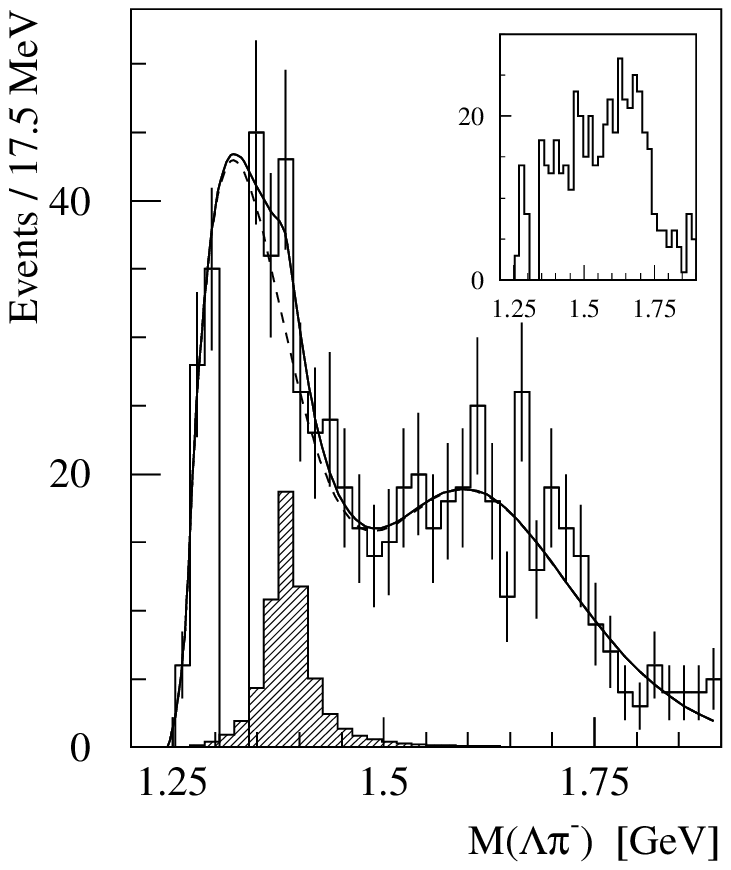}
\caption{%
    $M(\Lambda\pi^-)$ distribution for $\pT^2<0.001\,\mathrm{GeV}^2$,
    ignoring the $\Xi^-$ region in Fig.~\protect\ref{fig-bgr}.
    Example of a fit is shown by the solid line,
    which includes background for $\Sigma(1385)^-$ Coulomb
    production (dashed line) and possible signal.
    The latter is shown by the excess above the dashed line
    and by the hatched distribution, which corresponds
    to a derived $90\%$ CL upper limit for the radiative
    width of the $\Sigma(1385)^-$.
    The inset shows $M(\Lambda\pi^-)$ distribution
    for $0.002\,\mathrm{GeV}^2<\pT^2<0.006\,\mathrm{GeV}^2$.
}
\label{fig-fit}
\end{minipage}
\end{figure}

Beam $\Xi^-\to\Lambda\pi^-$ decays
are also selected by the above-described procedure.
Events in which $\Xi^-$ decays took place in the Pb target
constitute unavoidable background to the $\Lambda\pi^-$ final state.
To study this and all other\footnote{%
    For example,
    $\Sigma^-\to n\pi^-$ decay, with
    subsequent $n$ dissociation in matter into $p\pi^-$,
    mimics Reaction~(\protect\ref{re-main})
    when the $p\pi^-$ effective mass falls in the $\Lambda$ window
    and $\pT^2$ is small.
}
backgrounds caused by beam-particle decays,
two regions were defined for the primary vertex:
a target region ($-58\,\mathrm{cm}<Z_\mathrm{pri}<-52\,\mathrm{cm}$)
and sidebands upstream ($-63\,\mathrm{cm}<Z_\mathrm{pri}<-60\,\mathrm{cm}$)
and downstream ($-50\,\mathrm{cm}<Z_\mathrm{pri}<-47\,\mathrm{cm}$) of it.
Secondary vertex selection $Z_\mathrm{sec}>-45\,\mathrm{cm}$
provided equal $\Lambda$ decay probabilities for each region.
The effective mass distribution of the $\Lambda\pi^-$ systems
for events with very small values of the square
of the transverse momentum ($\pT^2<0.001\,\mathrm{GeV}^2$),
where most of the Coulomb production is concentrated,
is shown in Fig.~\ref{fig-bgr}.
Data for the primary vertex in the target region
are shown by the open histogram,
and data for sidebands are shown hatched.

A clear $\Xi^-$ peak is seen in the target region.
As expected, to within statistical accuracy,
the number of $\Xi^-$ decays in the sidebands is the same
as in the target region.
However, outside of the $\Xi^-$ peak,
there are almost no events in the sidebands.
Thus, with exception of
the $1.305\,\mathrm{GeV}$--$1.340\,\mathrm{GeV}$ region of $\Lambda\pi^-$ mass,
events with a primary vertex in the target region
are dominated by beam interactions, and
background from beam decays is negligible.

Although it is possible to subtract the background using the sideband regions,
the $\Xi^-$ peak is so dominant that the statistical error on the
difference exceeds any signal expected in that region.
That is why we decided simply to ignore
the region of $1.305\,\mathrm{GeV}$--$1.340\,\mathrm{GeV}$
in our analysis. The resulting distribution
(with a secondary vertex selection of $Z>-50\,\mathrm{cm}$,
to use the entire decay volume)
is shown as the histogram in Fig.~\ref{fig-fit}.
There is no clear evidence for $\Sigma(1385)^-$ production,
and we use the data to set an upper limit. This requires
an absolute normalization of the spectrum in Fig.~\ref{fig-fit},
and a description of the shape and magnitude of any expected signal.

The absolute cross section for the observed $\Lambda\pi^-$ spectrum
was determined by making use of the known $\Sigma^-/\pi^-$ ratio in the beam,
and by normalizing the data relative to similar diffractive
$\pi^+\pi^-\pi^-$ production on C and Pb nuclei,
and by then using the absolute $\pi^+\pi^-\pi^-$ production
cross section on C (see next paragraph).
As far as the trigger is concerned, 
there is no difference between the previously studied 
three-pion production on C and Pb targets~\cite{Molchanov:2001qk} 
and $\Lambda\pi^-$ production on the Pb target, 
provided that the $\Lambda$ decayed upstream of the IC counters.
Main trigger inefficiencies are due to pulse height selection
in the IC counters and accidental vetoes. These do not depend
on event topology. Thus, in such an analysis,
most trigger and all luminosity uncertainties cancel.
Remaining differences in efficiencies
of the trigger and of further event processing
are mainly due to different kinematics.
These are corrected using
a GEANT-based Monte-Carlo program~\cite{ge781-conf}.
In particular, for the $\Lambda\pi^-$ system
an angular distribution of $1+3\sin^2\theta\sin^2\phi$ is used,
as expected for electromagnetically produced $\Sigma(1385)^-$
in assumption of a purely magnetic-dipole
transition $\Sigma(1385)^-\to\Sigma^-\gamma$.
Due to high~($\sim70\%$, assuming the $\Lambda$ decayed in selected region)
geometrical acceptance of the SELEX apparatus to relevant processes,
systematical uncertainties of the relative normalization are negligible.

In this analysis, the diffractive production cross section
refers to the number of events observed
in the first diffractive exponential
of the $\pT^2$ distribution
for events in the $0.8\,\mathrm{GeV}<M(3\pi)<1.5\,\mathrm{GeV}$ mass range.
This was measured in special runs with a so-called ``beam'' trigger.
This trigger employed scintillation counters to define
beam particles and to reject halo, and required no information
from detectors downstream of the targets.
The measured value
$\sigma_\mathrm{diff}(\mathrm{C}) = 2.57\pm0.13\,\mathrm{mb}$
is based therefore on a completely unbiased set of interactions.
More detailed information on absolute normalization procedures
can be found in our previous publication~\cite{Molchanov:2001qk}.

The differential cross section
for the $\Lambda\pi^-$ Coulomb production by $\Sigma^-$
is given by the expression~\cite{halprin-1966-prim,faldt-1972-prim-b43}:
\begin{equation}
\frac{\mathrm{d} \sigma}{\mathrm{d} M \, \mathrm{d} q^2} =
\frac{2\alpha}{\pi}
Z^2
\frac{M}{M^2-m^2_{\Sigma^-}}
\,
\sigma_\gamma(M)
\,
\frac{q^2-q^2_\mathrm{min}}{q^4}
|F(q^2)|^2
\label{eq-primakoff_cs}
\end{equation}
where $\alpha$ is the fine structure constant,
$Z$~is the charge of the nucleus,
$M$~is the effective mass of the produced system,
$\sigma_\gamma(M)$~is the cross section
for the reaction $\gamma\Sigma^-\to\Lambda\pi^-$,
$q^2$~is the square of the momentum transfer,
$q^2_\mathrm{min}$~is its minimal value,
and
$F(q^2)$~is the nuclear form factor.
For the case of photoproduction through the $\Sigma(1385)^-$ resonance
(designated in the formulae as $\Sigma^*$),
the cross section $\sigma_\gamma(M)$
is given by the expression~\cite{Jackson:1964zd}:
\begin{equation}
\sigma_\gamma(M) =
8 \pi
\frac{M^2}{(M^2-m^2_{\Sigma^-})^2}
\,
\frac{2J_{\Sigma^*}+1}{2J_{\Sigma^-}+1}
\,
\frac{m_{\Sigma^*}^2\Gamma(\Sigma^-\gamma)\Gamma(\Lambda\pi^-)}%
{(M^2-m_{\Sigma^*}^2)^2 + m_{\Sigma^*}^2\Gamma^2_\mathrm{tot}}
\label{eq-photoproduction_cs}
\end{equation}
where $J$~stands for spin and $\Gamma$ for partial
(and $\Gamma_\mathrm{tot}$ for total) width.

The Coulomb form factor $F(q^2)$ in Eq.~(\ref{eq-primakoff_cs})
accounts for the nuclear charge distribution,
initial and final state absorption,
as well as the Coulomb phase.
It was calculated in the framework
of the optical model as developed in~Ref~\cite{bemporad-1973-prim}.
This model requires as input the
total $\Sigma^-$-nucleon cross section~$\sigma$,
and the ratio of real to imaginary parts
of the forward scattering amplitude~($\rho'$),
at the appropriate beam energy.
We used the cross section $\sigma=37.0\,\mathrm{mb}$
determined in the SELEX experiment,
and the extrapolated value of $\rho'=0.04$~\cite{Dersch:1999zg}.
The impact of these parameters on $F(q^2)$
at our energy and $q^2$ range is quite minimal.

At high beam energy, $q^2_\mathrm{min}$
can be approximated by:
\begin{equation}
q^2_\mathrm{min}
\approx 
(M^2-m^2_{\Sigma^-})^2/(4P^2_\mathrm{beam})
\end{equation}
For $M\approx1.385\,\mathrm{GeV}$, $q^2_\mathrm{min}$ is
$\approx3{\cdot}10^{-7}\,\mathrm{GeV}^2$.
Cross section~(\ref{eq-primakoff_cs}) peaks at
$q^2=2q^2_\mathrm{min}$, and then falls approximately as $1/q^2$.
Thus, the measured transverse momentum distribution
is determined mainly by experimental resolution,
which was obtained from studies of beam $\Xi^-\to\Lambda\pi^-$ decays.
In particular, since there is no momentum transfer in such decay,
the measured distribution directly reflects the resolution.
This was found to be Gaussian, with a standard deviation
varying from 16.3 to $17.8\,\mathrm{MeV}$,
depending on data set and cutoff criteria.
Equation~(\ref{eq-primakoff_cs}) was integrated numerically
to extract a cross section for $\pT^2<0.001\,\mathrm{GeV}^2$,
to be used for comparison with data.

When describing Primakoff production of the $\Sigma(1385)^-$ resonance,
its mass, total width, and branching to $\Lambda\pi^-$
were fixed to their known PDG values~\cite{Hagiwara:2002fs}.
The only fitted parameter was the radiative decay width. 
Background was described by two smooth functions,
vanishing at the $\Lambda\pi^-$ threshold.
The first function was
\begin{equation}
f_1(M) = P_1 \, (M-M_\mathrm{thr})^{P_2}
       \, \exp\left[{-{P_3}(M-M_\mathrm{thr})}\right]
\end{equation}
where $M_\mathrm{thr}$ was threshold mass, and $P_i$ were free parameters.
This function qualitatively described mass spectrum
in the region $M<1.5\,\mathrm{GeV}$.
The second function, capable of describing the entire mass spectrum, was
\begin{equation}
\begin{array}{rl}
f_2(M) = & P_1 \, (M-M_\mathrm{thr})^{P_2}
       \, \exp\left[{-{P_3}(M-M_\mathrm{thr})}\right] \\
       + & P_4 \, (M-M_\mathrm{thr})^{P_5}
       \, \exp\left[{-{P_6}(M-M_\mathrm{thr})}\right] \\
\end{array}
\end{equation}
We made fits using the whole data set, and using a reduced data set
where charm targets were absent, and, in particular, $\pT$ resolution
was better. There are arguments in favour of using for final analysis
each of these data sets, so we quote an average.
The difference is minor anycase.
Results of the fits and derived upper limits
are summarized in Table~\ref{tab-fits}.
An example of a fit
(function $f_2$, first line in Table~\ref{tab-fits})
is shown in Fig.~\ref{fig-fit}.
The fits showed no significant changes for small
variations in selection criteria and histogram binning.
However, significant uncertainty is associated with background description.
When giving an upper limit, there is
no conventional way of dealing with such type of uncertainty.
It was taken into account by using the worst case of function~$f_2$.

\begin{table}[!h]
\begin{center}
\begin{tabular}{|l|r|r|r|r|}
\hline
    Data set &
        \multicolumn{1}{c|}{$\Gamma^{(1)}_\mathrm{fit}$} &
            $\Gamma^{(1)}_{90\%}$ &
                \multicolumn{1}{c|}{$\Gamma^{(2)}_\mathrm{fit}$} &
                    $\Gamma^{(2)}_{90\%}$ \\
\hline
    All data &
        $0.37\pm3.79$ & 6.5 &
            $3.66\pm4.36$ & 9.8 \\
\hline
    Without charm targets &
        $-1.99\pm4.42$ & 6.2 &
            $1.32\pm5.08$ & 9.2 \\
\hline
\end{tabular}
\end{center}
\caption{Results of the fits for $\Gamma(\Sigma^-\gamma)$
         and derived upper limits (everything in keV).}
\label{tab-fits}
\end{table}

We quote an upper limit at the $90\%$ confidence level,
which includes all sources of uncertainty, as:
\begin{equation}
\label{eq-limit}
\Gamma\left[\Sigma(1385)^-\to\Sigma^-\gamma\right] < 9.5\,\mathrm{keV}
\end{equation}
Our result is dominated by statistical uncertainty
and uncertainty, associated with background description.
Other systematics uncertainties include
absolute normalization~($5\%$),
transverse momentum resolution~($1.8\%$),
accuracy of $F(q^2)$~($1\%$),
and uncertainties in the PDG parameters of the $\Sigma(1385)^-$ resonant
mass~($1\%$),
width~($5\%$),
and branching to $\Lambda\pi^-$~($2\%$).
Added in quadrature, these do not exceed~$8\%$.
The hatched distribution in Fig.~\ref{fig-fit} shows
what the signal spectrum would look like, given the
limit in~Eq.~(\ref{eq-limit}).

It is interesting to investigate what mechanism
is responsible for producing the $\Lambda\pi^-$ system
shown in Fig.~\ref{fig-fit}.
To do this, a $\Lambda\pi^-$ mass distribution
was studied for the $0.002\,\mathrm{GeV}^2<\pT^2<0.006\,\mathrm{GeV}^2$ range
(see the inset in Fig.~3),
which is almost free of Coulomb contribution,
and which is expected to be dominated by diffractive production.
The enhancement in Fig.~\ref{fig-fit}
in the region of $M\lesssim1.45\,\mathrm{GeV}$
is not present at higher transverse momenta.
This suggests an electromagnetic origin.
To check this, the $\pT^2$ distributions were studied
for mass regions below and above~$1.45\,\mathrm{GeV}$.
Both distributions showed strong forward peaking behavior,
which could be described by falling exponential $\exp(-b\pT^2)$.
For $M>1.45\,\mathrm{GeV}$ we obtained the slope
$b\simeq400\,\mathrm{GeV}^{-2}$,
a value typical for diffractive production on the Pb nucleus,
but for $M<1.45\,\mathrm{GeV}$, we obtained a slope
$b\simeq1000\pm100\,\mathrm{GeV}^{-2}$.
This is too steep to be explained by diffraction,
but quite reasonable assuming a significant Coulomb contribution.
Thus we conclude that $M\lesssim1.45\,\mathrm{GeV}$ region
is dominated by electromagnetic production.
Using Eq.~(\ref{eq-primakoff_cs}),
it is then possible to derive a cross section
for the reaction
\begin{equation}
\gamma\Sigma^-\to\Lambda\pi^-
\label{re-???}
\end{equation}
Available data do not allow us to separate
Coulomb and strong production mechanisms very precisely.
Also, there is an uncertainty in $\Lambda\pi^-$ detection efficiency,
since angular distribution of that system is not given by theory.
We estimate the cross section of Reaction~(\ref{re-???}),
including any possible $\Sigma(1385)^-$ resonant production,
to be
\begin{equation}
\sigma[\gamma\Sigma^-\to\Lambda\pi^-]
\Bigl|_{\sqrt{s}\approx1.385\,\mathrm{GeV}}
= 56\pm12\pm11\,\mu\mathrm{b}
\label{eq-sigma-measured}
\end{equation}
Factors, contributing to systematical uncertainty, include
absolute normalization~($5\%$),
transverse momentum resolution~($1.8\%$),
strong production~($10\%$),
and $\Lambda\pi^-$ detection efficiency~($15\%$).

The fact that electromagnetic production
dominates at low effective masses~$M$ is not surprising,
given the factor $M/(M^2-m^2_{\Sigma^-})$
in the cross section of~Eq.~(\ref{eq-primakoff_cs}).

\section{Discussion}

The upper limit of $9.5\,\mathrm{keV}$ at $90\%$ confidence level,
obtained for $\Gamma[\Sigma(1385)^-\to\Sigma^-\gamma]$,
is the best measurement to date, and improves by a factor of 2.5
the only previous limit of $24\,\mathrm{keV}$~\cite{Arik:1977mt}.
The current limit falls just at the edge
of the range of theoretical predictions~\cite{%
Lipkin:1973rw,Lipkin:1992vw,Akhiezer-eng,Darewych:1983yw,%
Hackman:1978am,Schat:1995mt,Abada:1996db,Haberichter:1997cp,%
Butler:1993pn,Napsuciale:1997ny,Leinweber:1993pv,Sahoo:1995kg,%
Wagner:1998bu,Wagner:2000ii}.

As far as we know, the result in Eq.~(\ref{eq-sigma-measured}) is the first
experimental measurement of photoproduction on~$\Sigma^-$.
We are not aware of any theoretical calculations for this value.
In view of this result,
it is interesting to discuss prospects for a new
$\Sigma(1385)^-\to\Sigma^-\gamma$ measurement.
Using Eq.~(\ref{eq-photoproduction_cs}),
the dependence between the cross section for resonance photoproduction
at the $\Sigma(1385)^-$ pole and the radiative width of this resonance
can be expressed numerically as
$\sigma_\gamma = 3.5\,\mu\mathrm{b}/\mathrm{keV}\times\Gamma_\mathrm{rad}$.
For $\Gamma_\mathrm{rad}$ in the predicted range,
the resonant cross section is small
compared to the total of~$56\,\mu\mathrm{b}$.
Thus, a measurement of the radiative width
will require a more sensitive experiment
by an order of magnitude if $\Gamma_\mathrm{rad}\approx4\,\mathrm{keV}$,
or by two orders of magnitude if $\Gamma_\mathrm{rad}\approx1\,\mathrm{keV}$.

Finally, we comment on a previous measurement~\cite{Arik:1977mt}.
In that experiment, the upper limit of $24\,\mathrm{keV}$
was based on 2 events observed in the $\Lambda\pi^-$ spectrum
in the $1.375\,\mathrm{GeV}<M<1.405\,\mathrm{GeV}$ mass range,
and correspond to at most 5 events at $90\%$~CL~\cite{Arik:1977mt}.
We have a factor of ${\approx}30$ more $\Lambda\pi^-$ events
in that mass range, while our upper limit is only $2.5$ times lower.
This is because an upper limit is determined mainly
by the statistical uncertainties of the electromagnetic
nonresonant component in $\Lambda\pi^-$ production.
Using our value in Eq.~(\ref{eq-sigma-measured})
for the $\Lambda\pi^-$ photoproduction cross section,
we predict $4.0\pm1.2$ events from Coulomb production
in the spectrum of Ref.~\cite{Arik:1977mt}.
Of course, there should also be a contribution from strong production.
Still, two detected events seems to be a credible fluctuation
down from an extrapolation of our result.

\section*{Acknowledgements}

%
%
The authors are indebted to the staff
of the Fermi National Accelerator Laboratory,
and for invaluable technical support from the staffs of collaborating
institutions.
This project was supported in part by Bundesministerium f\"ur Bildung, 
Wissenschaft, Forschung und Technologie, Consejo Nacional de 
Ciencia y Tecnolog\'{\i}a {\nobreak (CONACyT)},
Conselho Nacional de Desenvolvimento Cient\'{\i}fico e Tecnol\'ogico,
Fondo de Apoyo a la Investigaci\'on (UASLP),
Funda\c{c}\~ao de Amparo \`a Pesquisa do Estado de S\~ao Paulo (FAPESP),
the Israel Science Foundation founded by the Israel Academy of Sciences and 
Humanities, Istituto Nazionale di Fisica Nucleare (INFN),
the International Science Foundation (ISF),
the National Science Foundation (Phy \#9602178),
NATO (grant CR6.941058-1360/94),
the Russian Academy of Science,
the Russian Ministry of Science and Technology,
the Turkish Scientific and Technological Research Board (T\"{U}B\.ITAK),
the U.S. Department of Energy (DOE grant DE-FG02-91ER40664 and DOE contract
number DE-AC02-76CHO3000), and
the U.S.-Israel Binational Science Foundation (BSF).


\end{document}